\documentstyle[12pt]{article}
\begin{document}
\def\thebibliography#1{\section*{REFERENCES\markboth
 {REFERENCES}{REFERENCES}}\list
 {[\arabic{enumi}]}{\settowidth\labelwidth{[#1]}\leftmargin\labelwidth
 \advance\leftmargin\labelsep
 \usecounter{enumi}}
 \def\newblock{\hskip .11em plus .33em minus -.07em}
 \sloppy
 \sfcode`\.=1000\relax}
\let\endthebibliography=\endlist

\hoffset = -1truecm
\voffset = -2truecm


\title{\large\bf
Soft-QCD Effects In $B \Rightarrow \gamma \gamma$ Decay:
Quark Level Form Factors  
}
\author{
{\normalsize\bf
A.N.Mitra \thanks{e.mail: (1) ganmitra@nde.vsnl.net.in ;
(2) anmitra@csec.ernet.in}
}\\
\normalsize 244 Tagore Park, Delhi-110009, India.} 
\date{5 October 1999}
\maketitle

\begin{abstract}
        Soft-QCD effects of quark-level form factors in $\gamma\gamma$ decay 
of $B_s$-mesons via $D{\bar D}$ intermediate states, suggested by Ellis et al, 
are examined in the "Salpeter" model [reinterpreted in terms of the Markov-
Yukawa Transversality Principle (MYTP)] when formulated  on a covariant 
light front (null-plane). The gluonic kernel in the infrared regime which 
generates the constituent mass via standard $DB{\chi}S$, is calibrated to 
both $q{\bar q}$ and $qqq$ spectra, meson mass splittings, pion form factor, 
and other parameters.  With this check, an exact evaluation of the  
$D{\bar D}\gamma$ vertex form factor $F(\alpha)$ (normalized to $F(0)=1$), 
gives a multiplicative effect of $F^2(\alpha)$ on the $B_s \Rightarrow 
\gamma\gamma$ amplitude via $D$-meson triangle loop, where $\alpha$ = 
$(p^2+M^2)/M^2$ is the off-shellness parameter of the exchanged $D$-meson of 
mass $M$ and 4-momentum $p_\mu$, and is a variables of the (internal) loop 
integration. Unfortunately, the  $F^2(\alpha)$ effect decreases the loop 
amplitude by a factor of $30$ w.r.t. the point hadron value, resulting in a 
reduction of 3 orders of magnitude in long distance (hadronic) contributions 
to the $B_s \Rightarrow \gamma\gamma$ decay rate, thus greatly impairing the 
visibility of such modes.  

\end{abstract} 
  
\newpage

\section{Introduction: Hard VS Soft-QCD In B-Decays}

Rare B-decays carry signatures of important features of SM (Standard Model), 
such as FCNC (flavour changing neutral current), CKM (Cabibbo-Kobayashi-
Maskawa) in the electro-weak sector; and beyond SM, such as 2-Higgs doublet 
model, minimum SUSY extensions of SM, etc.  Of these decays, particular 
attention has been paid to $B_s \Rightarrow$ $\gamma\gamma$ decay [1-5], 
which (despite its small branching ratio) has a very clear signal where two 
monochromatic photons of high energy are produced back-to-back in the rest 
frame of $B_s$. Its present experimental limit on the branching ratio is [4] 
${\cal B} < 1.48 \times 10^{-4}$, but its sensitivity to hadronic acclerators 
(HERA,Tevatron and LHC), as well as to the planned $e^+e^-$ $B$-factories at 
KEK and SLAC, is likely to improve this limit significantly in the near 
future. This necessitates accurate theoretical estimates from different 
angles. Now within SM, the lowest order short distance contributions (box
diagrams, etc) give only ${\cal B} \approx 3.8 \times 10^{-7}$ [5], which is 
not much enhanced by Hard-QCD corrections [2]. This has led to mechanisms 
beyond SM [3a,3b]. On the other hand, even within SM, other options like a
$long-distance$ (hadronic) mechanism has been proposed for this process via 
$D^*{\bar D}^*$ intermediate states [5], giving big enhancements vis-a-vis 
short distance mechanisms with and without Hard-QCD corrections [1-2]. Such 
enhancements call for a closer scrutiny of this mechanism [5], in terms of
"Soft-QCD" corrections (to distinguish it from "Hard-QCD" corrections [2]),
and this is the purpose of this paper.     
\par
   Now Hard-QCD (leading log) corrections to short-distance effects are
compactly achieved by isolating a subsystem/process like $b->s\gamma\gamma$ 
[2]. More elegant methods for Hard-QCD corrections have been developed for 
heavy meson form factors [6a] under heavy-quark symmetry [6b]. On the other
hand, long distance contributions [5], where "Soft-QCD" effects play the 
dominant role, are expressed by the $D$-meson loop diagrams, fig.1a,  
where the $D{\bar D}\gamma$ vertex itself is not a point vertex, but an
extended structure consisting of a quark triangle loop with two hadron-quark, 
plus one quark-photon, vertices as shown in fig.1b. Thus the effect of 
Soft-QCD corrections boils down to the evaluation of this triangle loop, 
giving rise to a form factor $F(\alpha)$, where $\alpha$ is the off-shellness
parameter $1+p^2/M^2$ for the exchanged $D$-meson of momentum $p_\mu$ and mass
$M$ in fig.1a. Inserting $F(\alpha)$ at either vertex of fig.1a gives 
a weighting factor of $F^2$ in the $B_s \rightarrow \gamma\gamma$ amplitude 
via the $D$-meson triangle loop, where $\alpha$ may be conveniently taken
as one of the loop variables with its permissible range determined by the
kinematics of the decay amplitude.  In this respect, the effective 
$B_sD{\bar D}$ vertex [5] which constitutes the `source' of the $D{\bar D}$ 
annihilation, is a common background which  may be taken over directly from 
[5]. Further, it should be adequate to consider for simplicity a $D$-meson 
instead of $D^*$-meson triangle loop for the 2-photon annihilation, since the 
loop-integral structure is similar for both cases. 

\subsection{Theoretical Basis for Soft-QCD Corrections}  

Now, taking account of the state of the Soft-QCD art today, there are not 
many candidates for a satisfactory calculation of the quantity $F(\alpha)$
in a closed form, with the desired off-shell properties. Among the leading
candidates which have received a good deal of attention in the literature 
are heavy particle symmetry [6], QCD-SR [7] and chiral perturbation theory 
[8]. However the machinery offered by these do not seem to provide a closed 
non-perturbative form for the form factor $F(\alpha)$, when looked upon as a   
triangle loop integral (fig 1b) in terms of hadron-quark vertices. While 
heavy particle symmetry [6] is ideally suited to Hard-QCD effects, the actual 
formulations of the other two [7-8], [QCD-SR with its heavy reliance on the 
successive `twist' terms in an Wilson expansion; and chiral perturbation 
theory with its emphasis on expansions of the effective Lagrangians in the 
momenta], seem to lack the `closed form approach' that is so vital for  
ensuring a fully analytic form for $F(\alpha)$. [An analytic structure in
turn is essential for continuation from one physical region to another].   
\par
        With the hadron-quark vertex function as the ideal laboratory for 
testing Soft-QCD ideas, a more promising candidate for a closed form
approach is a BSE-SDE framework [9], wherein the starting point is a 4-fermion 
Lagrangian (current quarks) interacting via an effective gluon propagator in 
the non-perturbative regime [10]. This framework produces the mass function 
through the non-trivial solution of the SDE [10] as a concrete realization
of dynamical breaking of chiral symmetry ($DB{\chi}S$), in the sense of a 
generalized NJL mechanism [11] with space-time extended interaction. The mass 
function $m(p)$ in turn produces the bulk of the constituent mass in the
low momentum limit, via Politzer additivity [12]. We propose to use this
framework with the extra ingredient of a 3D support to this 4-fermion
interaction, offered by the Markov-Yukawa Transversality Principle (MYTP)
[13], whose relative lack of familiarity in the contemporary literature
warrants a short background which links it with the Salpeter equation [14].     

\subsection{MYTP As Covariant Basis For Salpeter Equation}
 
	The Salpeter model [14] which is based on the so-called adiabatic 
(instantaneous) approximation, results from a 3D reduction of the full BSE. 
Due to its non-covariant nature, its applications (although extensive) had 
been generally limited to topics like heavy quarkonia, with limited 
relativistic overtones. However its precise meaning in terms of a physical
principle got revealed by the work of the Pervushin Group (Dubna) [15] who 
showed that the Markov-Yukawa Transversality Principle (MYTP) [13] which 
decrees that the pairwise interaction be a function of the relative 4-
momentum ${\hat q}$ transverse to the 4-momentum $P$ of the composite, leads 
{\it exactly} to a 3D (Salpeter-like) form. Strictly speaking, such 
interactions which amount to a 3D support to the BSE kernel, are non-local, 
but reduce to the Salpeter form [14] in the rest frame of the composite, 
thus giving this equation [14] a fresh meaning within the MYTP context. 
\par 
	Now MYTP also possesses an interesting $dual$ property, viz., the
facility of an $exact$ $reconstruction$ of the 4D BSE through a reversal of 
steps from the 3D BSE form, as found subsequently by the Delhi Group [16a]. 
This twin property of MYTP, viz., an "exact interconnection"  between the 3D
and 4D forms of the BSE, gives the Salpeter equation a firm theoretical basis, 
for not only does MYTP reduce to it in the adiabatic limit, but has a wider 
(Lorentz-covariant) domain of applicability. (Indeed, this twin property 
of the Salpeter equation [14] had all along been present, but obscured from 
public view due to its non-covariant form !).  Further, the standard SDE-BSE 
framework [10]  characterized by the non-perturbative gluon propagator 
is easily adapted [16b] to MYTP, so as to produce an MYTP-constrained 
mass functions $m({\hat p})$, as a calculational tool for vacuum properties 
(like condensates) [16b], in a closed form.
\par
	The interlinked 3D-4D BSE structure which stems from MYTP, gives rise 
to a 2-tier mode of applications, the 3D form for the spectra [17], while 
the 4D form yields the hadron-quark vertex as the basic building block for 
calculating the 4D quark loop integrals [18], on the lines of the 
"dynamical perturbation theory" of Pagels-Stokar [19] (no criss-cross 
gluon lines inside the quark loops). These twin applications  serve as 
useful calibrations for the parameters of the model. More importantly, 
its  "exact" structure is ideally suited for imparting a crucial 
analyticity property to the function $F(\alpha)$, which carries the off-
shellness information down to the quark level of compositeness.      
\par
        Still another property of MYTP which stems from the characteristic 
$D \times \phi$ structure of the 4D vertex function [16a], ($D$ and $\phi$ 
being the 3D denominator fn and 3D wave fn respectively, which together 
satisfy the 3D BSE $D \phi$ = $\int K \phi$), concerns `quark confinement'. 
Thus in a 4D loop integral, such as fig.(1b), the $D$-fn causes a 
cancellation [16a] of the Landau-Cutkowsky singularities arising from 
integration over the loop variable, and thus prevents an otherwise free 
propagation of quarks inside the loops, a standard disease that usually 
plagues a quark loop calculation in terms of point vertex structures.        	
       
\subsection {MYTP w.r.t. The Covariant Null Plane}
 
	MYTP  still suffers from a serious problem, viz., a Lorentz 
mismatch of the participating vertex functions in loop integrals with 3 
or more quark lines, (e.g., fig.(1b)), leading to ill-defined integrals 
for the pion form factor and related applications [20], due to the 
appearance of time-like momenta in the 3D (gaussian) wave fns $\phi$. This 
shows up through a "complexity" in such amplitudes [20], where there is no 
physical reason for such behaviour. (On the other hand, 2-quark loops such 
as off-shell $\rho-\omega$ mixing [18a] and strong SU(2)-breaking for hadron 
mass splittings [18b]  just escape this pathology). This problem has 
recently found a simple solution through an obvious extension [21] of MYTP, 
so that the "transversality" w.r.t. the composite 4-momentum $P$ is 
defined covariantly on a null-plane whose orientation is specified by 2 dual 
4-vectors $n_\mu$, ${\tilde n}_\mu$, with $n^2={\tilde n}^2=0$; and  
$n.{\tilde n} =1$. The necessary details, making use of standard null-plane
technology [22] adapted to the covariant light-front, are given in [21]; it 
is shown that the reduced 3D BSE has a formally identical structure to that 
of the standard MYTP [13-16], so that the spectral predictions [17] remain 
unchanged. With this check as a "control" at the 3D level of spectroscopy , 
the application to 4D triangle loop integrals with the new vertex structures 
(albeit light-front orientation dependent), shows that they no longer suffer 
from the Lorentz mismatch disease, so that after the `pole' integrations 
over the time-like momenta, the loop integrals reduce to well-defined
3D forms, where the 3rd component depends covariantly on the null-plane 
orientation. In ref.[21], this new MYTP method is compared with covariant 
light-front (Null-plane) methods of Kadychevsky-Karmanov, and the 
Wilson Group's [23a-c], (collectively  reviewed in [24]). And as a basic 
application to the pion e.m. form factor,it is shown how a simple device of 
"Lorentz-completion"  yields a Lorentz-invariant structure with the correct 
asymptotic $1/k^2$ behaviour, when considered in a BSE-SDE framework [10] 
with a non- perturbative gluon propagator [16b]. The new MYTP method is 
ideally suited to the triangle loop integral for fig.(1b) which corresponds 
to the e.m. form factor of a $D$-meson, except that its domain of 
off-shellness differs from that of the pion form factor [21].        	       	 
\par
	In Sec.2, we collect some essential elements of the BSE model 
with 3D kernel support [15-16] in accordance with MYTP [13], as extended 
to the Covariant null-plane [21], on the lines indicated above. We write
down the full structure of the $B_s \rightarrow \gamma\gamma$ amplitude
$G_{\mu\nu}$ [5] in terms of the form factors $F(\alpha)$ of the two 
composite $D{\bar D}{\gamma}$ vertices involved in the $D$-meson triangle
loop [5]. In Sect 3, we sketch the derivation of $F(\alpha)$ in an exact
analytic fashion, and use this result in Sect 4 to evaluate $G_{\mu\nu}$ 
analytically over the entire kinematically allowed range of $\alpha$, 
resulting in a $30$ times reduced amplitude over the point hadron value 
[5], hence about a $1000$-fold reduction in the corresponding decay rate. 
Sect 5 summarises our findings, and notes the significance of this result 
vis-a-vis the observability of such radiative decay modes [5].
                                                   
\section{MYTP-Based BSE Model: 3D-4D Interlinkage}

In this Section we collect some essential ingredients of the MYTP [13]-
governed Bethe-Salpeter framework in preparation for the calculation of
the e.m. form factor of a hadron with unequal quark masses in the notation 
and phase convention of ref.[16], but adapted to the covariant null-plane 
(CNPA for short) [21]. To emphasize the close similarity of  the old
Covariant Instantaneity Ansatz (CIA) [16a], and the new CNPA [21] forms,
especially the exact 3D-4D interconnection among the respective wave
functions, we start with the principal ingredient, namely  the hadron-
quark vertex function which has the structure [16, 21]:                
\setcounter{equation}{0}
\renewcommand{\theequation}{2.\arabic{equation}}                
\begin{equation}
\gamma_D\Gamma({\hat q}) = N_n (P) D_n ({\hat q}) \phi({\hat q}) 
\gamma_D /(2\pi)^{5/2}
\end{equation}
Here ${\hat q}_{\mu} $ can be given a common meaning to cover both the 
CIA [16] and CNPA [21] situations: In CIA [16], it is the relative momentum
transverse to the hadronic 4-momentum $P_\mu$, i.e., ${\hat q}$ $=$ 
$q-q.P P/P^2$; while in CNPA [21] it is transverse to the null-plane $n\mu$,
i.e., ${\hat q}$ $=$ $q-q.{\tilde n}+n.qP.{\tilde n}n/n.P$ [21], where  
$n_\mu$ and ${\tilde n}_\mu$ are the null-plane 4-direction and its dual
respectively, normalized to $n^2={\tilde n}^2=0$ and $n.{\tilde n}=1$ [21].
In the standard $\pm$ notation [22], ${\hat q}$ translates exactly to 
to a 3-vector, viz., $q_\perp, q_3$ where $q_3=Mq_+/P_+$, with $P^2 = -M^2$ 
on the hadron mass shell. For off-shell hadron propagation (as is pertinent
for the problem on hand) on the other hand, $M^2$ should be replaced by 
$M^2(1-\alpha)$, where $\alpha$ = $(P^2+M^2)/M^2$ measures off-shellness. 
(The general $n_\mu$ dependence [21] keeps track of formal covariance, but for 
calculational purposes it is simpler (and faster) to use the old-fashioned 
notation [22]). Note that the $q_-$ component does {\it not} appear in 
${\hat q}$.  $\gamma_D$ is a Dirac matrix which equals $\gamma_5$ for 
pseudoscalar, $i\gamma_\mu$ for vector, etc hadrons [22c]; $D_n$ and $\phi$ 
are the 3D denominator and 3D wave function respectively [16]; $N_n(P)$ is the
BS normalizer. The explicit structures [16,21,22c] of the various symbols, 
generalized to off-shell hadron propagation, in the covariant NPA [21] are:  
\begin{equation}
q = {\hat m}_2p_1 - {\hat m}_1p_2 ; \quad 2{\hat m}_{1,2} = 1 \pm 
(m_1^2-m_2^2)/M^2;  
\end{equation}
\begin{equation}
D_n({\hat q}) = 2P.n [q_\perp^2 + M^2(1-\alpha)q_+^2/P_+^2 - 
\lambda(M^2(1-\alpha), m_1^2, m_2^2)/M^2(1-\alpha)]
\end{equation}
where $\lambda$ is the standard triangle function of its arguments. The
corresponding denominator function $D_n({\hat q}')$ is the same as (2.3)
above, except for the replacements $P,q \rightarrow P',q'$ and $\alpha=0$.     
$\phi = exp (-{\hat q}^2/2\beta^2)$ is the 3D wave function whose inverse 
range parameter $\beta$ is a dynamical function [21] of the basic constants 
of the (input) BS kernel [17]; and similarly for $\phi'$.   
\par
        We shall make use of this framework to calculate the quark level
form factor $F(\alpha)$ at each of the two $D{\bar D}\gamma$ vertices of
the hadron triangle (fig. 1a) for 2-photon decay of $B_s$, a la fig.1b. 
Here $\alpha$ is the off-shellness parameter for the exchanged $D_s$-meson 
in fig.1a which corresponds to fig.1 of ref.[5]. Inserting these two form 
factors in the corresponging amplitude $G_{\mu\nu}$ of ref.[5] for 
$B_s \rightarrow \gamma\gamma$, the latter in our momentum notation 
(fig.1a) and euclidean convention, reads [5]
\begin{eqnarray}
G_{\mu\nu} &=& e^2{\cal G}f \int {d^4p \over {(2\pi)^4}} [(K^2-P_1^2)f_+
+P_2^2 f_-] F^2(\alpha)   \\  \nonumber
           & & {{(4p_\mu p_\nu - \delta_{\mu\nu}) (p^2+M^2)} \over 
{(p^2+M^2)(p+k_1)^2+M^2) ((p-k_2)^2+M^2)}} + [1 \Rightarrow 2]
\end{eqnarray}
where $K=P_1+P_2=k_1+k_2$ is the total 4-momentum, and the
weak interaction parameters ${\cal G}, f, f_{\pm}$ are as defined in 
ref.[5]; $p$ is the 4-momentum of the exchanged $D_s$-meson of mass $M$, 
so that $p^2+M^2=M^2 \alpha$. The effect of the contact $\gamma\gamma$ 
interaction in the hadron loop, viz., $G_{\mu\nu}^{(3)}$ of ref.[5], may be 
recognized in this `master expression', via the term proportional to 
$\delta_{\mu\nu}$ in the numerator of the integrand on the right, which has 
been simplified by dropping some terms that vanish on contracting with the 
photon polarizations: $\epsilon_1.k_1$ = $\epsilon_2.k_2$ = $0$.  
   
\begin{figure}[t]

\caption{}
\vspace{0.5in}

\begin{picture}(450,175)(-30,-10)
\put(35,10) {(a) $B_s \Rightarrow \gamma\gamma$} 
\put(230,10) {(b) Triangle loop for vertex `O' in (a)}

\multiput(25,110)(-3,7){7}{\line(1,0){2.8}}
\multiput(25,110)(-3,7){7}{\line(0,1){7}}
\multiput(95,110)(3,7){7}{\line(-1,0){2.8}}
\multiput(95,110)(3,7){7}{\line(0,1){7}}

\put (25,110){\line(1,0){70}}
\put (25,111.5){\line(1,0){70}}

\put (55,118.5){\line(3,-2){10}}
\put (55,105){\line(3,2){10}}

\put (230,60){\line(1,0){40}}
\put (230,63){\line(1,0){40}}

\put (370,60){\line(1,0){40}}
\put (370,63){\line(1,0){40}}

\put (245,68){\line(3,-2){10}}
\put (245,54.5){\line(3,2){10}}

\put(245,48){\makebox(0,0){$P_1=P'$}}
\put(390,48){\makebox(0,0){$P$}}
\put(290,95){\makebox(0,0){$p'_1$}}

\put (385,68){\line(3,-2){10}}
\put (385,54.5){\line(3,2){10}}

\put (270,68){\line(3,-2){10}}
\put (270,54.5){\line(3,2){10}}
\put (270,68){\line(0,-1){13}}

\put (360,61.5){\line(3,-2){10}}
\put (360,61.5){\line(3,2){10}}
\put (370,68){\line(0,-1){13}}

\multiput(320,110)(0,20){3}{\line(1,1){10}}
\multiput(330,120)(0,20){3}{\line(-1,1){10}}
\put (314,130){\line(5,2){10}} 
\put (324,134){\line(5,-2){10}}

\put(344,149){\makebox(0,0){$k_1=k$}}

\put(320,52){\makebox(0,0){$-p_2$}}

\put(26,112){\makebox(0,0){\large O}}
\put(95,112){\makebox(0,0){\large O}}

\put(60,121){\makebox(0,0){$p=P_1-k_1$}}

\put(60,90){\makebox(0,0){\large ${\nabla}$}}

\put(58.75,57){\rule{1mm}{10mm}}

\put(60,44){\makebox(0,0){$B_s(K)$}}

\put(60,76){\line(-1,-2){5}}
\put(60,76){\line(1,-2){5}}

\put(22,94){\makebox(0,0){$D(P_1)$}}
\put(10,170){\makebox(0,0){$\gamma(k_1)$}}

\put(98,94){\makebox(0,0){${\bar D}(P_2)$}}
\put(112,170){\makebox(0,0){$\gamma(k_2)$}}

\put (25,110){\line(5,-3){30}}
\put (28,110){\line(5,-3){28}}
\put (95,110){\line(-5,-3){30}}
\put (92,110){\line(-5,-3){28}}

\put (93.5,130){\line(3,2){10}}
\put (104,137){\line(1,-4){3}}

\put (16,137){\line(3,-2){10}}
\put (11,127){\line(1,2){5}}

\put (320,110){\vector(1,-1){25}} 
\put (344.5,85.2){\line(1,-1){20}}

\put (275,64.7){\vector(1,1){25}}
\put (299.5,89.5){\line(1,1){20}}

\put(352,95){\makebox(0,0){$p_1$}}

\put (360,61.5){\vector(-1,0){45}}
\put (335,61.5){\line(-1,0){55}}

\end{picture}
\end{figure}

\par
        Our next task is to write down the expression for  $F(\alpha)$, 
when the exchanged hadron $p_\mu$  (fig.1a) is off-shell. To that end we 
temporarily relabel $p_\mu$ as $P_\mu$, with $P^2=-M^2 (1-\alpha)$,
and $P_1=P'$ with $P'^2=-M^2$, so as to conform to fig.1b, and the following 
expression for $F(\alpha)$ emerges in a fairly standard fashion [21, 22c]  
\begin{equation}
2e{\bar P}_\mu i F(\alpha) = (2\pi)^{-4} e N_n(P)N_n(P') \int d^4p_2 
D_nD_n'\phi \phi' 4 T_\mu /[\Delta_1 \Delta_1' \Delta_2] + 
[`1' \Rightarrow `2']
\end{equation}
\begin{equation}
4T_\mu = {\Delta_1 \Delta_1' \Delta_2} \times 
Tr[\gamma_5S_F(p_1)i\gamma_\mu S_F(p_1') \gamma_5 S_F(-p_2)]   
\end{equation}
which simplifies to
\begin{equation}
T_\mu = {\bar P_\mu}[\Delta_2 +\Delta_1/2 +\Delta_1'/2 - x_2 k^2/2
-(1-{\hat m}_2)({\delta m}^2 +P^2/2 -M^2/2)]
\end{equation}
where $\Delta_i = p_i^2 + m_i^2$, $i=1,2$, are the inverse propagators, and 
the substitution ${\hat m}_2 \approx p_2.{bar P}/{\bar P}^2$, as the 
fraction of $p_2$ in the direction of ${\bar P}$, has been made in the last 
term of $T_\mu$, eq.(2.7). The BS normalizers $N_n(P,P')$ which correspond 
to both hadrons on-shell ($\alpha = 0$) and photon 4-momentum $k_\mu =0$, 
which are expressible, by Lorentz covariance, in terms of the invariant 
hadron normalizers $N_H$ as $N_n(P,P')= N_H (M/P_+, M/P_+')$, may be 
inferred from (2.5-7) [21], and the condition $F(0) \equiv 1$ (see eq.(3.7)
below, for an explicit formula), so that the value of $F(\alpha)$ away 
from $F(0)$ is a direct measure of the form factor effect on the point 
$D{\bar D}\gamma$ vertex [5]. The kinematical range of $\alpha$ which can 
be related to the cosine of the scattering angle in the subprocess 
$D{\bar D} \rightarrow \gamma\gamma$, fig.1a, works out as
\begin{equation}
\alpha_{min,max} = {{M_s^2 \mp M_s \sqrt {M_s^2 - 4M^2}} \over {2M^2}}
\end{equation}
where the energetics are controlled by the mass $M_s$ of the `source'
hadron $B_s$. Substituting for the respective masses [25], the limits
work out as $1.20 < \alpha < 6.26$.  
\par
    We now turn to the evaluation of $F(\alpha)$, eq.(2.5), in Sect.3, and 
use this quantity for evaluating $G_{\mu\nu}$, eq.(2.4), in Sect.4 next. 
   
\section{Exact Evaluation of $F(\alpha)$}
\setcounter{equation}{0}
\renewcommand{\theequation}{3.\arabic{equation}}

        The evaluation of $F(\alpha)$, eq.(2.5), follows closely a recent
calculation of the pion form factor [21], except that one of the hadrons
($P_\mu$) is now off shell, and the photon is on shell, while in the pion
form factor case [21], it was the other way around. To recount the main
steps, note first that since $q_-$ is absent from $\phi({\hat q})$, all 
the `pole' singularities in the $q_-$-plane are contained only in the 3
quark propagators in eq.(2.5). The detailed techniques of pole residues
may be found in [22c], but we summarize the formulae [21]:  
\begin{equation}
\oint {dp_{2-} \over 2} {1 \over \Delta_2}[{1 \over \Delta_1}; {1 \over 
\Delta_1'};{1 \over {\Delta_1 \Delta_1'}}] = 2i\pi [{1 \over D_+}; 
{1 \over D_+'};{{2p_{2+}} \over {D_+D_+'}}] 
\end{equation}
in the standard $\pm$-component notation [21,22] for the $(D_n, D_n')$-
functions associated with the hadronic 4-momenta $P_\mu$ (off-shell) and 
$P_\mu'$ (on-shell) respectively. Next, we collect some definitions and 
the results of some simplification after integration over $dp_{2-}$    
\begin{equation}
q_++ q_+' = 2{\bar q}_+ ; \quad z_2 \equiv {\bar q}_+/{\bar P}_+ ; \quad 
p_{2+} = {\hat m}_2P_+ -{\bar q}_+ ;
\quad      
\end{equation}            
\begin{equation}
\phi\phi' = exp[-q_{\perp}^2/\beta^2 - W^2 z_2^2/\beta^2]; 
\quad  W^2 \equiv M^2(1- \alpha/2) 
\end{equation}
And the trace factor ${Tr}_+$$\equiv$$T_\mu/{\bar P}_\mu$ in Eq.(2.6) is
\begin{equation}
{Tr}_+ = 2{\bar P}_+ [q_{\perp}^2+ W^2 (z_2^2-1/4) 
+(m_1^2+m_2^2/2 -{{(\Delta m^2)^2 W^2} \over {4M^4 (1-\alpha)}}]
+ 2p_{2+} {\hat m}_1 [W^2 - {\delta m}^2] 
\end{equation} 
where ${\Delta m^2}= m_1^2-m_2^2$. Collecting all the pieces of $F(\alpha)$ 
from eqs.(3.1-4) and putting them in (2.5) yields a simple quadrature:
\begin{equation}
F(\alpha) = 2N_H^2 {M^2 \over {P_+P_+'}} \int d^2 q_{\perp} {\bar P}_+dz_2 
{Tr}_+ \phi\phi' 
\end{equation}
the result of whose integration gives the explicit formula 
\begin{equation}
F(\alpha) = {{4N_H^2 M^2} \over {(2\pi)^3 W}} 
(\pi\beta^2)^{3/2} [{{3\beta^2} \over 2} + {(m_1^2+m_2^2) \over 2}-W^2/4 
+M^2 \sigma^2 \theta +2{\hat m}_1{\hat m}_2 (W^2-{\delta m}^2)]
\end{equation}   
where the simplification $P_+=P_+'={\bar P}_+$ has been employed in view
of the on-shellness of the emitted photon ($k^2=0$). $\sigma$ is defined 
as ${\Delta m^2}/2M^2$, and $\theta$=$(1-\alpha/2)/(\alpha-1)$. 
From this the normalizer $N_H$ is inferred in the limit $\alpha=0$ as 
\begin{equation}
N_H^{-2} = {{4M} \over {(2\pi)^3}} (\pi \beta^2)^{3/2}  
[3\beta^2/2 +(m_1^2+m_2^2)/2 -M^2/4 -M^2 \sigma^2 + 
2{\hat m}_1{\hat m}_2 (M^2-{\delta m}^2)]
\end{equation}  
which is symmetrical in $(m_1, m_2)$ and ensures $F(0) \equiv 1$. 
\par
        Since form factors like $F(\alpha)$ are a typical feature of 
hadron compositeness, additional singularities should be expected in the 
integrand for $G_{\mu\nu}$, vis-a-vis a point-hadron description [5] where 
its only singularity is the pole $\alpha=0$, corresponding to the 
on-shell value of the hadron propagator. Indeed (3.6) shows a branch point 
at $\alpha=2$, but its effect on $G_{\mu\nu}$ is harmless; see Sect.4 below.     
 
\section{Exact Evaluation Of $G_{\mu\nu}$}
\setcounter{equation}{0}
\renewcommand{\theequation}{4.\arabic{equation}}

Our next task is to evaluate the decay amplitude (2.4) by inserting in it 
the value (3.6-7) for $F(\alpha)$, with the kinematical range for $\alpha$ 
as $ 1.2 < \alpha < 6.26$. The ratio of this amplitude to one with a point
$D{\bar D}\gamma$ vertex ($F(\alpha)=1$) will measure the effect of hadronic
compositeness on the `long-distance' (soft-QCD) comtribution to this 
decay process [5]. The strategy is thus to evaluate (2.4) with and without 
the $F^2(\alpha)$ factor under the same approximation as in ref.[5], viz.,
replacing the two propagators $[P_{1,2}^2+M^2]^{-1}$, on the rhs of (2.4) 
by their on-shell values $i\pi\delta(P_{1,2}^2+M^2)$. Next, explicit gauge 
invariance is achieved by ensuring the proportionality of $G_{\mu\nu}$ to 
$Q^2 \delta_{\mu\nu} - 2Q_\mu Q_\nu$, where $2Q = k_1-k_2$, since it vanishes
on contraction separately with $k_{1\mu}$,  or with $k_{2\nu}$, and using the 
results $k_1.Q =2Q^2$, etc. To extract this combination from the 
integrand at the earliest, the $d^4p$ integration in (2.4) may be arranged
in accordance with the resolution of the 4-vector $p_\mu$ in 4 mutually  
perpendicular directions: $p_\mu$ = $p_{\perp \mu} +p.QQ_\mu/Q^2$ 
$+p.KK_\mu/K^2$, where $K=k_1+k_2$, $Q=(k_1-k_2)/2$, and $Q.K=0$. 
\par
	 Now the 4D measure may be expressed as 
\begin{equation}
d^4p = d^2p_\perp d(p.Q) d(p.K)/[M_s\sqrt{Q^2}]
\end{equation} 
and in association with the absorptive parts (delta-fns) of the two 
$D$-propagators, the 4D measure gives the net result
\begin{equation}
d^4p \pi^2 \delta(p^2+2p.Q+M^2) \delta(p.K)/2 = 
d^2p_\perp \pi^2/[4M_s\sqrt{Q^2}]; \quad p.K=0; \quad 2p.Q=-M^2 \alpha
\end{equation}      
Thus effectively $p_\mu$ and $Q_\mu$ are 3-vectors, and the off-shell
parameter $\alpha$ corresponds to the angle between them. And the 2D
measure $d^2p$, on integration w.r.t. the azimuthal angle (not involved in 
the denominators), becomes $\pi d({p_\perp}^2)$, where ${p_\perp}^2$ is 
entirely expressible in terms of $\alpha$:
\begin{equation}
{p_\perp}^2 = M^2(\alpha-1) - M^4 \alpha^2/4Q^2; \quad
d^2p_\perp=M^2(1-M^2 \alpha/{2Q^2})d\alpha
\end{equation}
which reduces the 4D integration to a simple quadrature in $\alpha$ only. 
\par
	Since the azimuthal angle is involved only in the factor 
$4p_\mu p_\nu$ in (2.4), its integration gives 
\begin{equation}
<4p_\mu p_\nu> = 2 \Theta_{\mu\nu} {p_\perp}^2 +4x^2 Q_\mu Q_\nu; 
\quad x= p.Q/Q^2; \quad \Theta_{\mu\nu}= \delta_{\mu\nu}-2Q_\mu Q_\nu /Q^2
\end{equation}
To ensure explicit gauge invariance, we employ a pedagogical method which
checks exactly with the final result of ref.[5] in the point hadron limit. 
This consists in `regularizing' a certain (apparently non-gauge invariant) 
portion from  $[\delta_{\mu\nu}(p^2+M^2)-4<p_\mu p_\nu>]$, following the 
classical Bethe-Schweber treatment of vacuum polarization in QED [26]. 
Of this the piece $\Theta_{\mu\nu}$ is gauge invariant by itself, but
the `unwanted' piece $[\delta_{\mu\nu}(p^2+M^2)-2x^2 Q^2]$ must be isolated
and `regularized'. The net result which gives $G_{\mu\nu}$ as a product
$\Theta_{\mu\nu}$ $\times$ ${\cal A}$, is a simple quadrature in $\alpha$:
\begin{equation}
{\cal A}= C \int d\alpha {{F^2(\alpha)} \over \alpha} (1-\lambda \alpha/2) 
(\lambda \alpha^2/2+1-\alpha); \quad \lambda = M^2/Q^2     
\end{equation}
where the constant $C$ represents the effect of the electroweak factors 
as employed in ref.[5]:
$$ C = M^2{{{\cal G}fe^2 [(M_s^2-M^2)f_+ - M^2 f_-]} \over {16\pi}}$$
and  $F(\alpha)$ is given by (3.6-7). And except for a slightly different 
integration strategy to accommodate the form factor $F(\alpha)$, this
formula checks with ref.[5] in the point hadron limit. The constant $C$
may now be dropped as we are interested only in checking the relative
effect of $F(\alpha)$ on the point-hadron result of ref.[5]. 
\par
	The integration w.r.t. $\alpha$ in $1.20 < \alpha < 6.26$ is
elementary even after the inclusion of $F(\alpha)$. In the point hadron
limit [5], its value is     	
\begin{equation}
{\cal A}_0 = J_0(6.26)-J_0(1.20)= -0.4824; \quad 
J_0(x) = \ln x -(1+\lambda/2) x +\lambda x^2/2 - \lambda^2 x^3/12
\end{equation}  
In the composite hadron case, eq.(3.6) suggests a branch point singularity
at $\alpha=2$ which is the limit up to which the (gaussian) integral for
$F(\alpha)$ is defined. This gives a simple pole structure for $F^2(\alpha)$ 
at $\alpha=2$; beyond this point $F^2(\alpha)$ may be defined by analytic 
continuation. [The singularity at $\alpha=1$ is outside its range of 
integration]. The integral (as a Principal value) after all substitutions is
\begin{equation}
{\cal A} =  \int_{1.20}^{6.26} {dx \over {x(1-x/2)}} (1-x+x^2\lambda/2)
[a-(1-x/2)(b+c/(1-x))]^2 = -0.01464    
\end{equation}
where $\alpha$ has been renamed as $x$ and
\begin{eqnarray}
aN &=& 3/2 \beta^2+(m_1^2+m_2^2)/2 -2{\hat m}_1{\hat m}_2{\delta m}^2 \\ \nonumber
bN &=& M^2/4 -2{\hat m}_1{\hat m}_2 M^2; \quad cN = M^2\sigma^2  \\  \nonumber
N  &=& 3/2\beta^2+m_2^2-(M^2-m_1^2+m_2^2)^2/{4M^2}+
2{\hat m}_1{\hat m}_2(M^2-{\delta m}^2)
\end{eqnarray}
The numerical values of these quantities were obtained by substitution 
from the spectroscopic values of their basic ingredients [17,21]:
\begin{equation}
a = +0/9446; \quad b = -0.2370; \quad c = +0.2882
\end{equation}
leading to the final value $(-0.01464)$ for ${\cal A}$, eq.(4.7). As a result
of this exercise, the ratio of the amplitudes for the composite (4.7) to the
point hadron limit (4.6) is $.03035$, giving a reduction in the decay rate
by the factor $\rho \approx 900$, i.e., by 3 orders of magnitude. 

\section{Resume And Conclusion}
	     
	We have tried to estimate the effect of quark compositeness on
the long distance contributions to $B_s \Rightarrow \gamma\gamma$ [5], 
within a BSE framework under the Markov-Yukawa Transversality Principle 
(MYTP) on the BSE kernel for $q{\bar q}$ interaction, which gives an exact
interconnection between the 3D and 4D forms of the BS amplitude. A major
advantage of this formalism (which is fully calibrated to the spectra [17]
and other observables [18,21]), vis-a-vis several other candidates [6-8] in
the field, is that it provides an exact analytic structure for the 
$D{\bar D}\gamma$ form factor in terms of the off-shell energy variable 
$\alpha$ of the exchanged hadron. Further, when folded into the $D$-triangle 
loop for the $B_s \rightarrow \gamma\gamma$ amplitude, it gives an analytic 
structure of the integral in this variable, and thus offers a basic
confidence in the reliability of the quark compositeness effect due to its
prior calibration to spectroscopy. The result for the decay rate is a 
reduction by three orders of magnitude over the point hadron value [5]. And 
although the calculation was made for the $D{\bar D}$ annihilation mode for 
simplicity, the mechanism is general enough to apply to the more pertinent 
$D^*{\bar D}^*$ mode as well, as it seems to be the more promising candidate 
for the measurability of such long distance modes of radiative $B_s$ decays 
in hadronic B-factories (HERA-B, CDF, D0, LHC-B), as and when available [5].              
\par
        I acknowledge useful comments from Prof S.R.Choudhury on this paper.

\end{document}